\documentclass[namedreferences]{SolarPhysics} 
\usepackage{epsfig}
\usepackage{psfig}
\usepackage{graphicx}
\usepackage{times}
\usepackage{color}
\newcommand{\arcsec}{\hbox{$^{\prime\prime}$}}
\newcommand{\ion}[2]{#1\,{\small #2}}
\newcommand{\todash}{\,--\,}

\begin{document}
\begin{article}
\begin{opening}

\title{Spectral Line Selection for HMI: A Comparison of \ion{Fe}{I} 6173 \AA~and \ion{Ni}{I} 6768 \AA}
\author{A.A. \surname{NORTON}$^{1,2}$,
	J. \surname{PIETARILA GRAHAM}$^{2}$, 
	R.K. \surname{ULRICH}$^{3}$,
	J. \surname{SCHOU}$^{4}$, 
	S. \surname{TOMCZYK}$^{2}$,
	Y. \surname{LIU}$^{4}$,
	B.W. \surname{LITES}$^{2}$,
	A. \surname{L\'OPEZ ARISTE}$^{2,5}$,
	R.I. \surname{BUSH}$^{4}$,
	H. \surname{SOCAS-NAVARRO}$^{2}$,
	P.H. \surname{SCHERRER}$^{4}$, }

\runningauthor{NORTON {\it ET AL.}}
\runningtitle{HMI Line Selection}

\institute{$^{1}$ National Solar Observatory, 950 N. Cherry Ave., Tucson, AZ 85719, USA
		\email{norton@nso.edu}\\
		$^{2}$HAO, National Center for Atmospheric Research, Boulder, CO 80301, USA\\
 		$^{3}$Department of Physics and Astronomy, University of California at Los Angeles, Los Angeles, CA 90095, USA\\
		$^{4}$HEPL, Stanford University, Stanford, CA 94305, USA\\
		$^{5}$THEMIS. CNRS UPS 853. La Laguna-38200. Spain\\
		}

\date{Received; accepted}

\begin{abstract}
We present a study of two spectral lines, \ion{Fe}{I} 6173 \AA~and \ion{Ni}{I} 6768 \AA, that were candidates to be used in the Helioseismic and Magnetic Imager (HMI) for observing Doppler velocity and the vector magnetic field.   The line profiles were studied using the Mt. Wilson Observatory, the Advanced Stokes Polarimeter and the Kitt Peak McMath telescope and one meter Fourier transform spectrometer atlas.  Both \ion{Fe}{I} and \ion{Ni}{I} profiles have clean continua and no blends that threaten instrument performance.   The \ion{Fe}{I} line is 2\% deeper, 15\% narrower and has  a 6\% smaller equivalent width than the \ion{Ni}{I} line.

The potential of each spectral line to recover pre-assigned solar conditions is tested using a least-squares minimization technique to fit Milne-Eddington models to tens of thousands of line profiles that have been sampled at five spectral positions across the line.   Overall, the \ion{Fe}{I} line has a better performance than the \ion{Ni}{I} line for vector magnetic field retrieval.   Specifically, the \ion{Fe}{I} line is able to determine field strength, longitudinal and transverse flux four times more accurately than the \ion{Ni}{I} line in active regions.    Inclination and azimuthal angles can be recovered to $\approx$2$^\circ$ above 600 Mx/cm$^2$ for \ion{Fe}{I} and above 1000 Mx/cm$^2$ for \ion{Ni}{I}.   Therefore, the \ion{Fe}{I} line better determines the magnetic field orientation in plage, whereas both lines provide good orientation determination in penumbra and umbra.  We selected the \ion{Fe}{I} spectral line for use in HMI due to its better performance for magnetic diagnostics while not sacrificing velocity information.  The one exception to the better performance of the \ion{Fe}{I} line is when high field strengths combine with high velocities to move the spectral line beyond effective sampling range.   The higher $g_{eff}$ of \ion{Fe}{I} means that its useful range of velocity values in regions of strong magnetic field is smaller than \ion{Ni}{I}. 

\end{abstract}

\end{opening}

\section{Introduction}
\label{intro}

The Helioseismic and Magnetic Imager (HMI) needs to accurately measure Doppler velocity and vector magnetic field with limited spectral information, sampling roughly half a dozen wavelengths across the spectral line.   HMI is one of a suite of instruments to be included on-board the Solar Dynamics Observatory (SDO) with launch currently scheduled for August of 2008.  The spacecraft is expected to have a large velocity range of $\pm$4000 m/s.  

The first requirement for choosing a spectral line is that it contain a clean continuum with no blends and no near-by lines.  Helioseismology requires that a spectral line be narrow and deep since a steep $dI/d\lambda$ ensures greater sensitivity to small Doppler shifts.  The \ion{Ni}{I} 6768 \AA~line satisfied these requirements and was selected as the spectral line for the Michelson Doppler Imager (MDI) and Global Oscillation Network Group (GONG) instruments.  Vector magnetic field measurements benefit from a high Land\'e factor, g$_{eff}$, and a simple Zeeman splitting geometry.  It is also important to minimize the number of blends that become apparent in sunspot umbrae where the lower temperatures allow for molecular absorption.  

An understanding of the center-to-limb variations of the selected spectral line is important because the accuracy of ``look-up" algorithms (such as used by MDI) are based on line depth and wing-slope parameters.  In addition, it is desirable for a line to be relatively insensitive to thermodynamic changes so that heights of formation do not drastically change due to moderate perturbations of temperature and density.  Of course, optimum instrumental transmission at the selected wavelength is also a requirement.  

The potential of each spectral line to recover pre-assigned solar conditions is tested using a least-squares minimization technique to fit Milne-Eddington models to tens of thousands of line profiles that have been sampled at five spectral positions across the line.  See Figure \ref{JGFILTERS} for an example of filter profiles applied at five positions across the \ion{Ni}{I} line.  Similar research was carried out by Graham {\it et al.}, (2002) to determine the minimum number of spectral filters necessary to recover the vector magnetic field parameters.   The work contained in this paper builds on that premise insofar as we examine the differences inherent in two spectral lines in their capacity to recover the vector magnetic field with filter-polarimetric measurements.  

\section{General Information of Spectral Lines}

Five papers  discuss the solar spectral \ion{Fe}{I} 6173 \AA~line (Stenflo and Lindegren, 1977; Auer {\it et al.}, 1977; Simmons and Blackwell, 1982; Solanki and Stenflo, 1985; Landi degl'Innocenti, 1982).  
Two papers detail the \ion{Ni}{I} 6768 \AA~line (Jones, 1989; Bruls, 1993).  The spectral line central wavelength, effective Land\'e factor and excitation potential are noted in these papers and can be found in the first three columns of Table 1.  Also found in Table 1 are the line depths and full-width half maxima values as seen in the disk center quiet sun Fourier transform spectrometer (FTS) atlas.  The FTS data were used because the system does not suffer from contamination by scattered light.  Line depth and width values for the spectral lines may vary slightly when observed with different optical systems.  

The heights of line formation are also found in the last two columns of Table 1.  They are estimated using Maltby-M umbral model (Maltby {\it et al.}, 1986)
and VAL-C model (Vernazza {\it et al.}, 1973, 1976, 1981) under non-LTE assumption. Stokes profiles were simulated using a non-LTE numerical radiative transfer code (Uitenbroek, 2001) based on the multilevel accelerated lambda iteration (MALI) formalism of Rybicki and Hummer (1991).   
All calculations were made in a one-dimensional plane-parallel geometry.  The model atoms used were a 52-level atom used for Fe calculations and a 25-level atom used for Ni.   The calculated results are listed in the last two columns of Table I with the first row being the VAL-C model height for core and continuum and the second row being the Maltby-M model height.   

\begin{table}[!ht]
\begin{center}
\label{tab1}
\caption{Parameters of \ion{Fe}{I} 6173 \AA~and \ion{Ni}{I} 6768 \AA}
\bigskip
\begin{tabular}{ccccccc} \hline \hline
Wavelength & g$_{eff}$& Excit. Pot.& Depth & FWHM & H (core) & H (cont.)\\ 
(\AA)& & (eV)& &(\AA) & (km)  & (km)\\
\hline
\ion{Fe}{I} 6173.34 & 2.499& 2.22& 0.66& 0.102 & 302  & 16\\
        &      &     &     &      & 269 & 21  \\
\ion{Ni}{I} 6767.68 & 1.426& 1.83& 0.64& 0.116 & 288  & 18\\
        &      &     &     &      & 291 & 26  \\
\hline \hline
\end{tabular}
\end{center}
\end{table}
The calculations of \ion{Ni}{I} height of formation agree well with other literature sources, but the \ion{Fe}{I} heights differ from Bruls (1993). The Bruls values for \ion{Fe}{I} are h = 238 (core) in the quiet Sun using the VAL-C model and h = 386 (core) in umbrae using the Maltby-M model.

\subsection{Line Profiles and Transmission in Sunspot Umbra and in Quiet Sun}

Line profiles of \ion{Fe}{I} 6173 \AA~and \ion{Ni}{I} 6768 \AA~in sunspot umbra 
and in quiet Sun scanned by the Kitt Peak McMath telescope and one meter Fourier 
transform spectrometer are available at \texttt{ftp://argo.tuc.noao.edu/pub/atlas}.  Due to page limitations, we can not reproduce this data and instead, discuss only the important features. 

The profiles of these lines in the quiet sun are relatively clean.  \ion{Ni}{I} has no nearby lines or blends.   \ion{Fe}{I} has a blend 0.6 \AA~away from its line center identified as \ion{La}{II}, which is at 6172.72 \AA.  This \ion{La}{II} blend is the transition of 3F3 \todash{} 3F2 where the lower level excitation potential is 0.13 eV and the $g_{eff}$ is 1.5. There are many Zeeman components (15-plet).  It does not simply split, it broadens as shown in the photographic sunspot atlas.  Simulations have been carried out to estimate the \ion{La}{II} blends influence on Doppler velocity measurements.   In a worst case scenario, results show this blend introduces systematic errors of ~1 or 2 m/s if its presence is not taken into account.  Of course, the HMI look-up algorithm for Doppler measurements will account for the existence and behavior of the \ion{La}{II} blend in order to minimize errors.

\begin{figure*}[htbp]
\centering{\epsfig{file=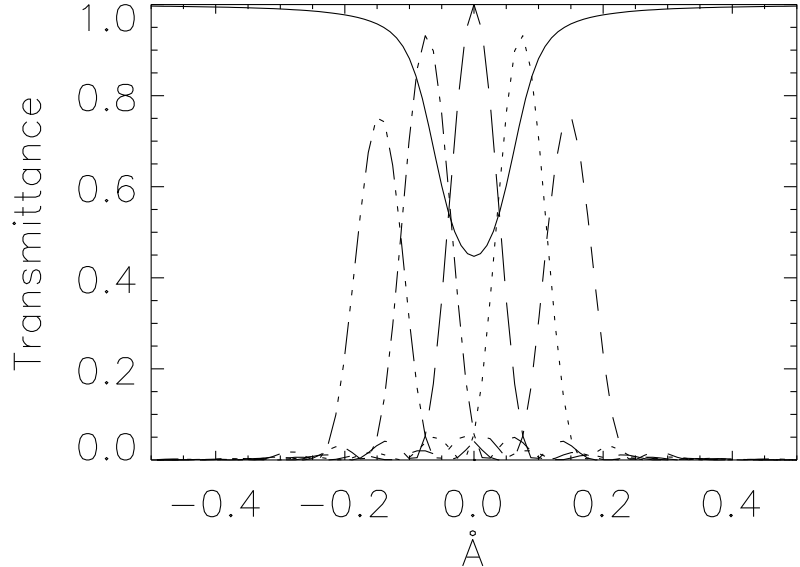,width=25pc}}
\caption{One possible configuration of the HMI filter set overlayed with the absorption line for \ion{Ni}{I} 6768~\AA~(solid line).  The central filter is 90~m\AA~FWHM and the spacing is 76~m\AA~between filter centers.}
\label{JGFILTERS}
\end{figure*}

Since HMI will make vector magnetic field observations, the behavior of these lines in umbrae is of interest.  It is important to realize that almost all umbral profiles are contaminated by molecular blends because the lower temperature of the sunspot allows for such spectral absorption not present in the quiet Sun spectra.  The \ion{Fe}{I} umbral profile shows an obvious blend in the blue wing. This blend is invisible in the quiet Sun spectra.  This blend is \ion{Eu}{II} at 6173.0 \AA~and there are other two blends near this line.  \ion{Ni}{I} observations in umbra have shown an obvious blend suggested to be TiO and two other blends nearby. 

\subsection{Trade-off Between Land\'e Factor and Velocity Range}
There is a trade-off between a spectral line with higher g$_{eff}$ that performs better for magnetic field diagnostics and a line with a lower g$_{eff}$ for which the velocity calculation algorithm performs better.   Having a line with a higher g$_{eff}$ means that the Zeeman splitting is greater and the useful velocity range is smaller in the presence of a strong magnetic field.  When the velocity algorithm no longer responds linearly to the Doppler shift due to one wing of the line profile having moved out of the spectral sampling range, this is called ``saturation".  (Note, this is distinctly different from the saturation experienced using traditional magnetograph algorithms when the Stokes $V$ lobes stop growing but begin separating with increasing field strength.)  

A shift in the line profile of 0.140 \AA~causes one wing of the profile to move out of sampling range.  (This value is slightly less than twice the filter spacing of 76 m\AA.  See Figure~\ref{JGFILTERS} for the configuration of HMI filters.)  We use the Doppler shift formula of $\delta\lambda=\lambda_{rest} \times v/c$ to determine that \ion{Fe}{I} will saturate at approximately 6.8 km/s while \ion{Ni}{I} will saturate at approximately 6.2 km/s due to the Doppler shift alone.   If the spectral line components are shifted solely due to the Zeeman effect, we use the formula $\delta\lambda = 4.67\times10^{-13} g_{eff} \lambda^{2} B$ where $\delta\lambda$ and $\lambda$ are in \AA, $g_{eff}$ is the effective Land\'e factor and $B$ is in gauss, to determine that a magnetic field of approximately 3.1 kGauss will cause \ion{Fe}{I} to saturate whereas a 4.4 kGauss will cause \ion{Ni}{I} to saturate.  Note that \ion{Fe}{I} has a g$_{eff}$ of 2.5 while \ion{Ni}{I} has a g$_{eff}$ of 1.4.  Of course, when the line is shifted due to both velocity and magnetic field, ``saturation" will occur at lower values.  

\begin{figure*}[]
\centering {\epsfig{file=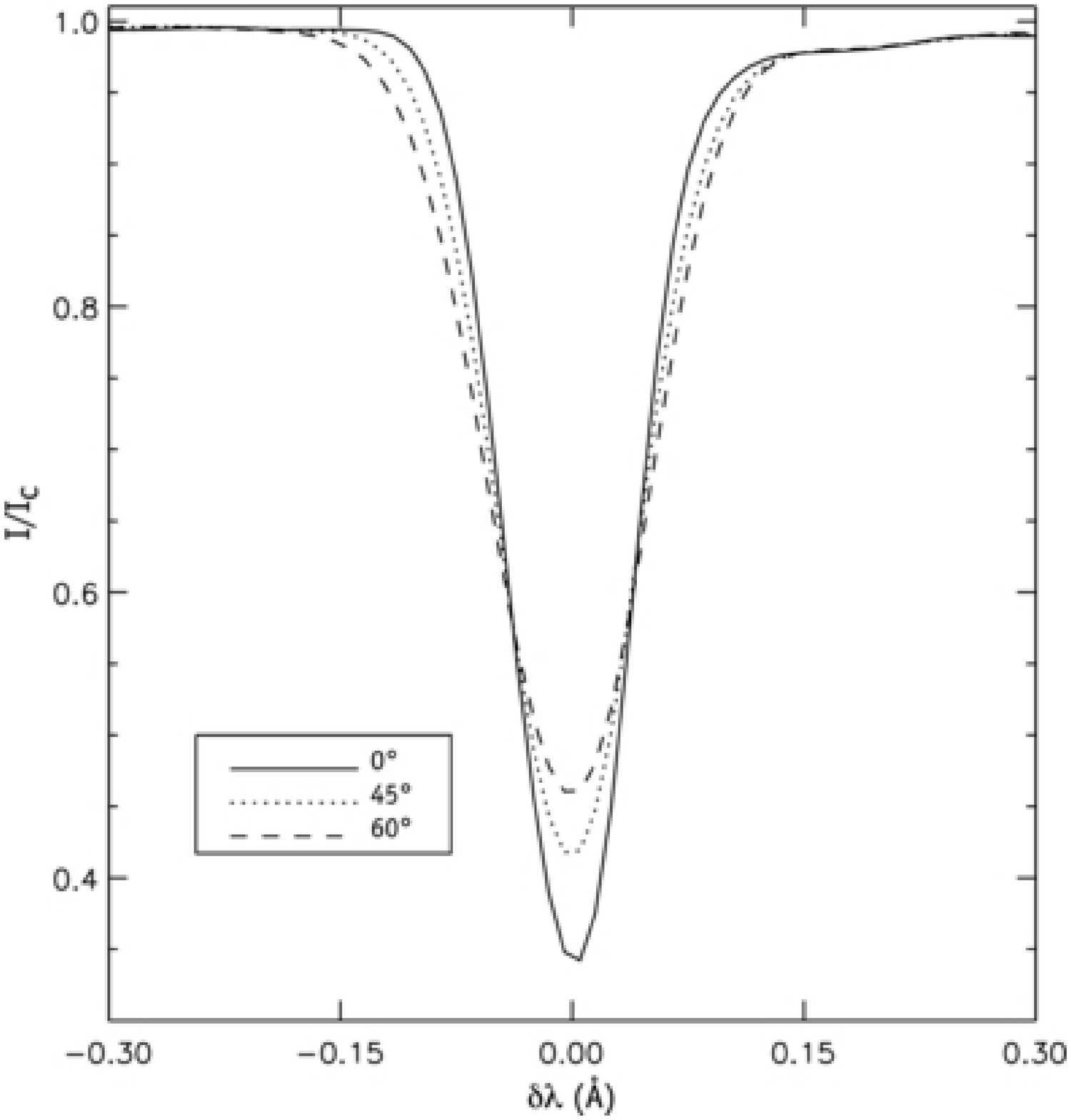,width=20pc}}
\centering {\epsfig{file=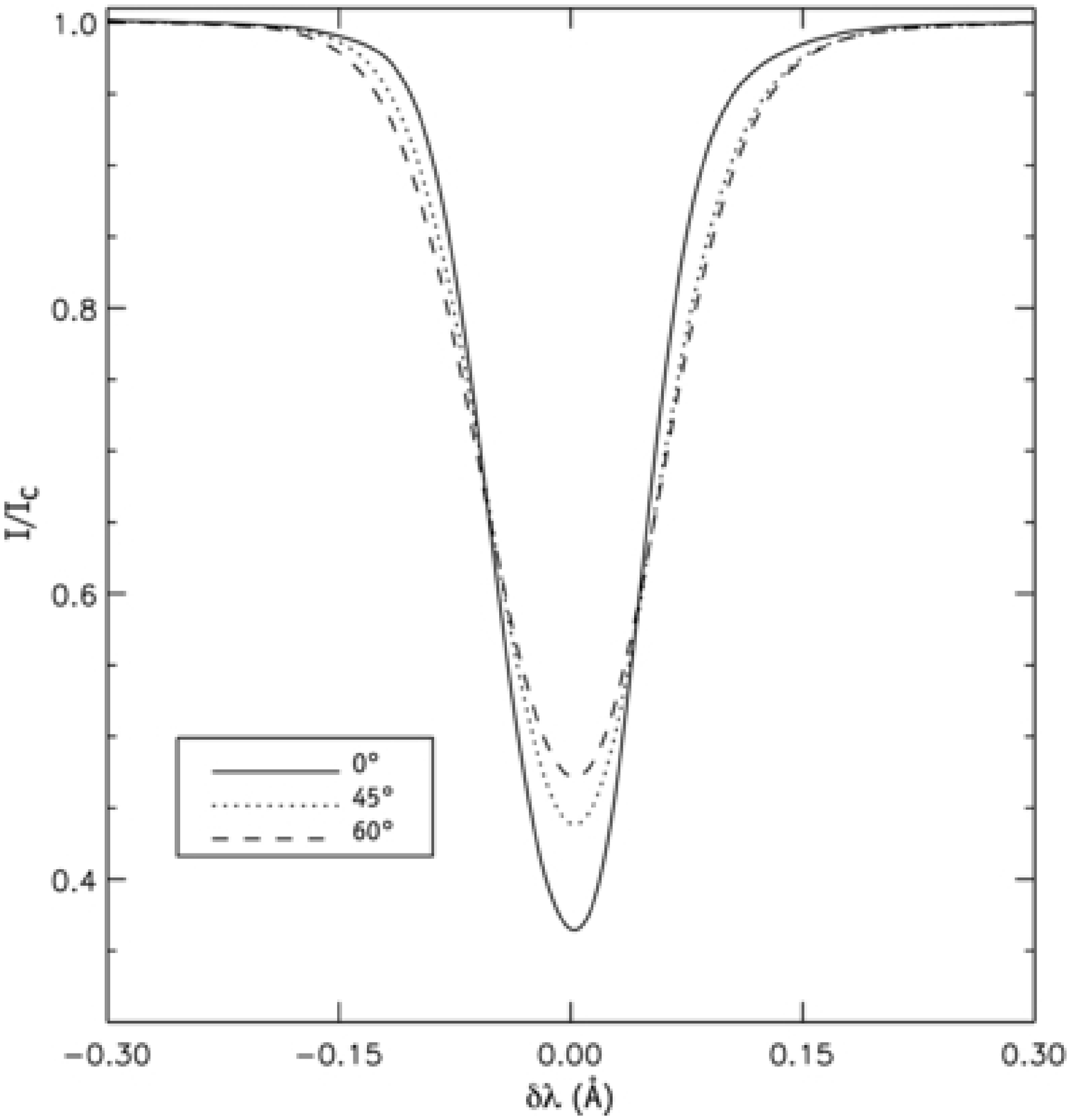,width=20pc}}
\caption{The line profiles of \ion{Fe}{I} 6173 \AA~(upper) and \ion{Ni}{I} 6768 \AA~(lower) obtained at Mt. Wilson for the quiet Sun regions. The profiles are shown for three center-to-limb angles:  0, 45 and 60 degrees.}
\label{6173w}
\end{figure*}

\begin{figure*}[]
\centering {\epsfig{file=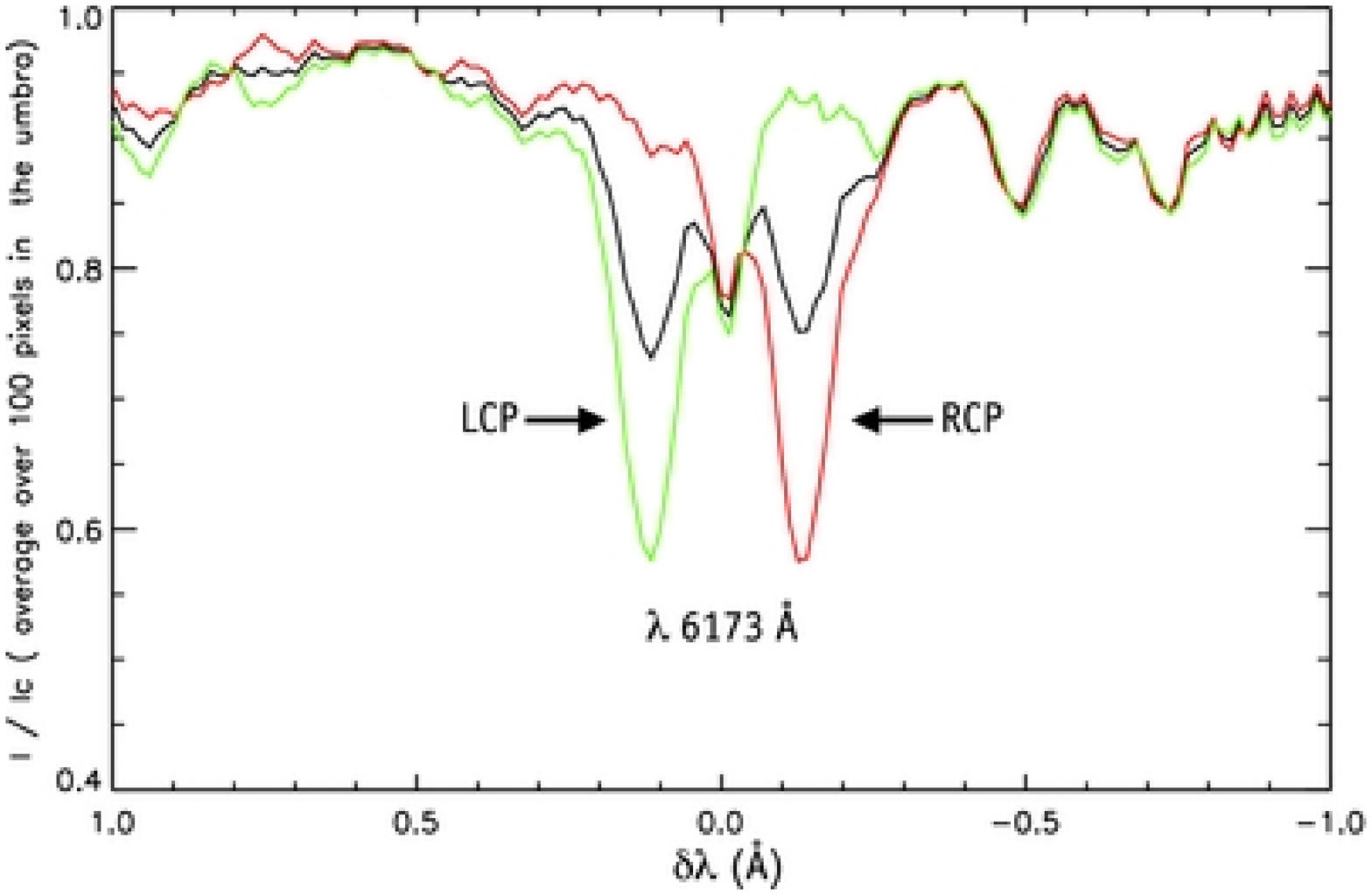,width=24pc}}
\centering {\epsfig{file=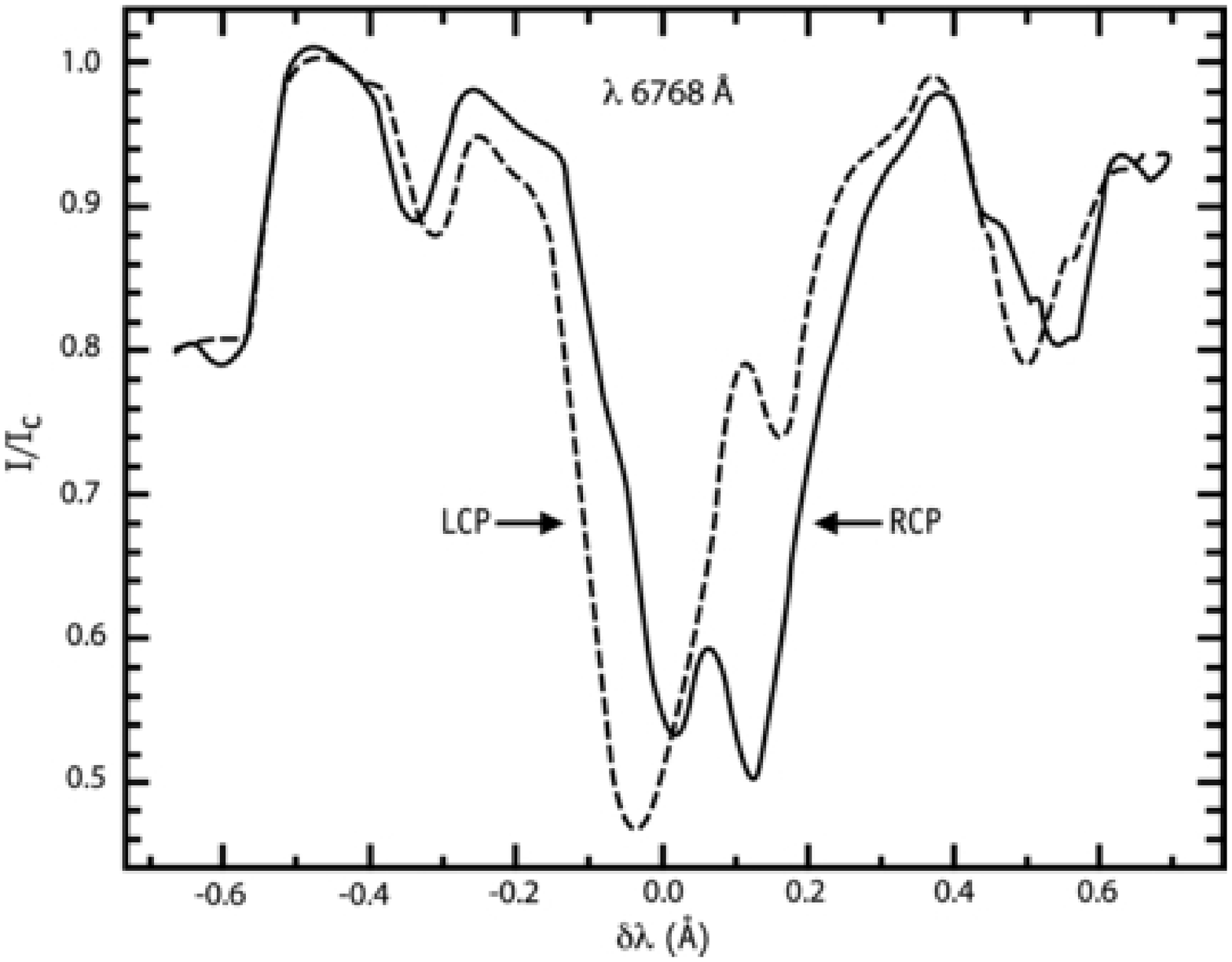,width=24pc}}
\caption{\ion{Fe}{I} 6173 \AA~(upper) observed by the ASP. LCP (green) and RCP (red) profiles are shown as well as Stokes $I$ (black).  The profiles shown here are an average over 100 pixels (pixel size is 0.6$\arcsec \times$ 0.37$\arcsec$) in the umbra.  \ion{Ni}{I} 6768 \AA~(lower) line in left and right circular polarizations observed at Mt. Wilson.}
\label{6173asp}
\end{figure*}

\section {Observations of \ion{Fe}I 6173 \AA~and Observations of \ion{Ni}{I} 6768 \AA~using Mt. Wilson Observatory and Advanced Stokes Polarimeter}

The magnetograph at Mt. Wilson Observatory has sufficient resolution and sensitivity to scan profiles of the spectral lines. The Mt.\ Wilson system for line profile measurements was described in its initial configuration
by Ulrich {\it et al.} (1991) and in its extended configuration by Ulrich {\it et al.} (2002).  The scanned profile of the line \ion{Ni}{I} in a sunspot umbra shows a pattern in satisfactory agreement with that obtained by the Kitt Peak McMath telescope and 1 meter Fourier transform spectrometer. The profile also displays all of the fine structure as shown in the Kitt Peak's spectrum. 

\ion{Fe}{I} line was scanned at Mt. Wilson for quiet sun regions on 2002 December 4; see the top panel in Figure~\ref{6173w}.    The Mt.\ Wilson system uses a moving stage that carries a bundle of fiber-optic image reformattors
with independent rectangular pickups in order to scan the spectrum.  The outputs from the pickups are measured
by ten independent photomultiplier tubes to form independent channels of intensity measurement.  In 1991 there were two active channels while in 2002/2003 there were ten active channels.  The spectral range from each pickup is offset from the next due to the geometry of the reformattor bundle so that a different range of spectrum is scanned by each of the channels.  The profiles used here are based on the restricted range of spectrum that is covered by all active channels.  The scanning was alternatively red to blue then blue to red taking a total of 20 minutes.  Thus all of these profiles are smeared by the five-minute oscillations. The resulting profiles are shown in the top panel of Figure~\ref{6173w} for three center-to-limb angles:  0, 45 and 60 degrees.  The line profiles of the \ion{Ni}{I} were obtained in 1991 at Mt. Wilson for the quiet Sun. The resulting profiles are shown in the bottom panel of Figure~\ref{6173w} for three center-to-limb angles; 0, 45 and 60 degrees.  The individual profiles for each sub-scan are not smeared and have a narrower profile.  These are a better representation of the profiles measured by the HMI filters since each exposure is short.  These profiles were not used because individually they are noisier than the time-averaged profiles and because the exposure duration represented by the time-averaged profiles more closely matches the exposure duration for the ASP.
 
Two properties of the Mt.\ Wilson system have degraded the observed line profiles:
\begin{enumerate}
\item The spectral resolution is about 25 milli-\AA ngstroms due to the finite size of
the slit and spectrograph.  This degradation is corrected by a single pass
deconvolution based on a laser profile in 1991 and based on a sodium cell
absorption profile in 2002/2003.
\item The Littrow spectrograph system has internal
scattered light which uniformly illuminates the entrance apertures of the spectral
sampling system.  Based on a comparison of the very deep Na D lines to observations
made with double pass spectrograph systems, we find that the scattered light
level is about 2\%\ of the continuum level.  This is an additive
offset that should be added to all the intensities.
\end{enumerate}
 
Stokes parameters $I$, $Q$, $U$, and $V$ were acquired by the Advanced Stokes Polarimeter (ASP) on 2002 March 9 with the \ion{Fe}{I} line observing in an active region umbra located at S3 W4. The line profiles of \ion{Fe}{I} in the umbra in circular polarizations are shown in top panel Figure~\ref{6173asp}. Also plotted in this Figure is the Stokes I in black, as a reference. The profiles are averaged over 100 pixels (pixel size is 0.6$\arcsec \times$ 0.37$\arcsec$) in this umbra.  The left and right-circular polarizations of \ion{Ni}{I} for a sunspot umbra were also obtained at Mt. Wilson, which are shown in lower panel of Figure~\ref{6173asp}. The two blends, the \ion{TiO} line and \ion{CaH} line, appear to be shifted for the left and right circular polarizations, suggesting that the effective Land\'e factors are not equal to zero.

\subsection {Spectral Line Parameters as a Function of Center-to-Limb Angle}

\begin{figure*}[ht]
\centering {\epsfig{file=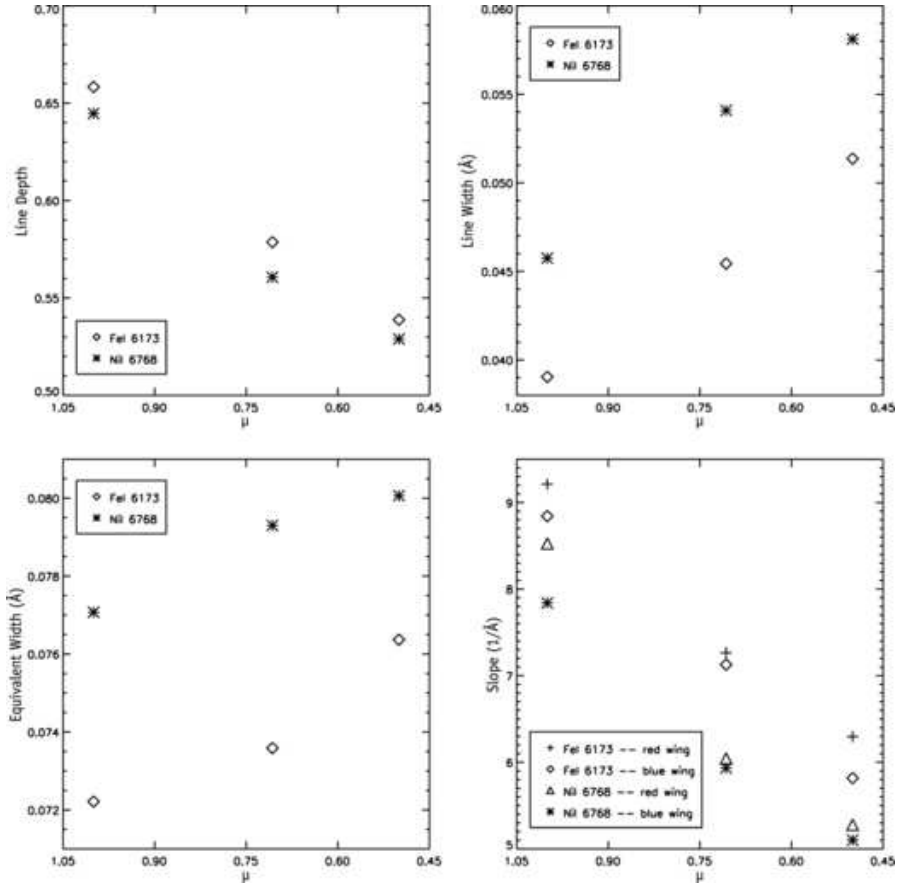,width=28pc}}
\caption{Center-to-limb variation of the lines \ion{Fe}{I} 6173 \AA~and \ion{Ni}{I} 6768 \AA~in quiet Sun observed at Mt. Wilson. The $x$-axis is the central meridian distance of the observations as in $\mu$=cos($\theta$).}
\label{cl6768}
\end{figure*}

The line depth, width, equivalent width, and slopes for the red and blue wings of the spectral lines \ion{Fe}{I} 6173 \AA~and \ion{Ni}{I} 6768 \AA~for the quiet Sun have been measured from Mt. Wilson data and can be seen in Figure~\ref{cl6768}. The line depth and width are defined here as the height and width of a Gaussian function which best fits the observed line profile. Since measurement for \ion{Fe}{I} was only made at three positions, \( 0^o, 45^o, 60^o \), in the quiet Sun, we used the \ion{Ni}{I} data at the same positions for a comparison. 

 Basically, the line depths for both lines linearly decrease with the cosine of the center-to-limb angle, the line widths linearly increase, and the equivalent widths slightly increase. The \ion{Fe}{I} line is generally 2\% deeper and 15\% narrower than that of \ion{Ni}{I}. The equivalent width of \ion{Fe}{I} is 6\% smaller than that of \ion{Ni}{I}. The slopes for the red wings of the two lines are generally greater than those for the blue wing; while the slope of \ion{Fe}{I} is greater than that of \ion{Ni}{I}. 

\subsection {Vector Magnetic Field Retrieval of \ion{Ni}{I} 6768 \AA~and \ion{Fe}{I} 6173 \AA~Using the \ion{Fe}{I} 6301/6302 \AA~Line Pair as a Reference}

Comparisons of ASP inverted data for the \ion{Ni}{I} and \ion{Fe}{I} lines show how well the lines perform for the purpose of vector magnetometry. ASP inversions are a least squares fitting based on the Milne-Eddington solution of the Unno-Rachkovsky equations of a plane-parallel magnetized radiative transfer of the Stokes line profiles. Normal ASP operation utilizes the \ion{Fe}{I} line pair of 6301 \AA~and 6302 \AA.  However, another spectral line can be observed simultaneously. On 2002 March 9, a map of an active region was made by scanning the ASP spectral slit across a sunspot near disk center (S3 W4) observing with the \ion{Fe}{I} 6302 \AA/6301 \AA~line pair and simultaneously scanning the \ion{Fe}{I} 6173 \AA~line.  On 2002 March 10, a map of a limbward active region (S9 E62) was made observing with the \ion{Fe}{I} 6302 \AA/6301 \AA~line pair  and \ion{Ni}{I} 6768 \AA~line.

\begin{figure*}
\centering {\epsfig{file=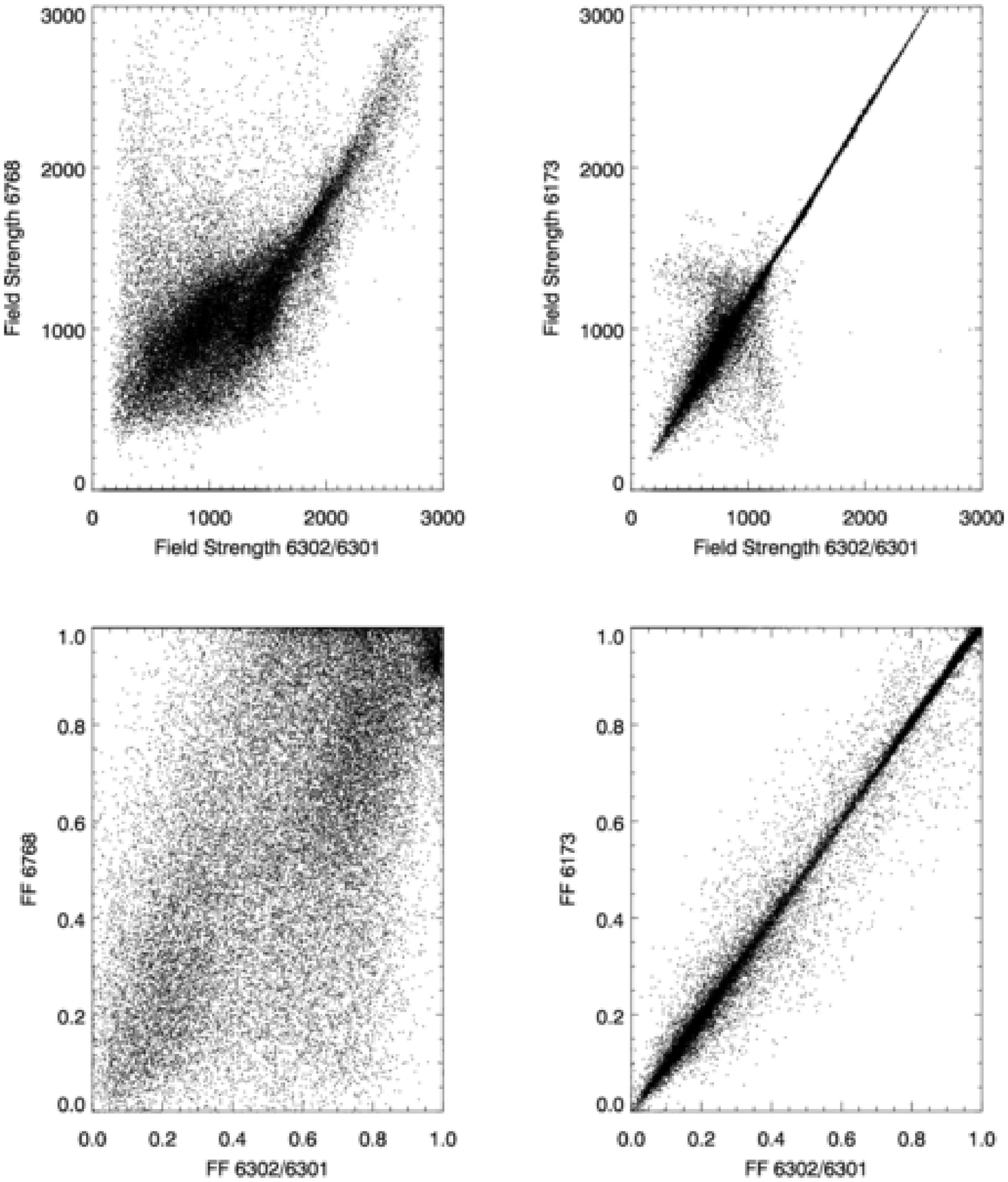,width=25pc}}
\centering {\epsfig{file=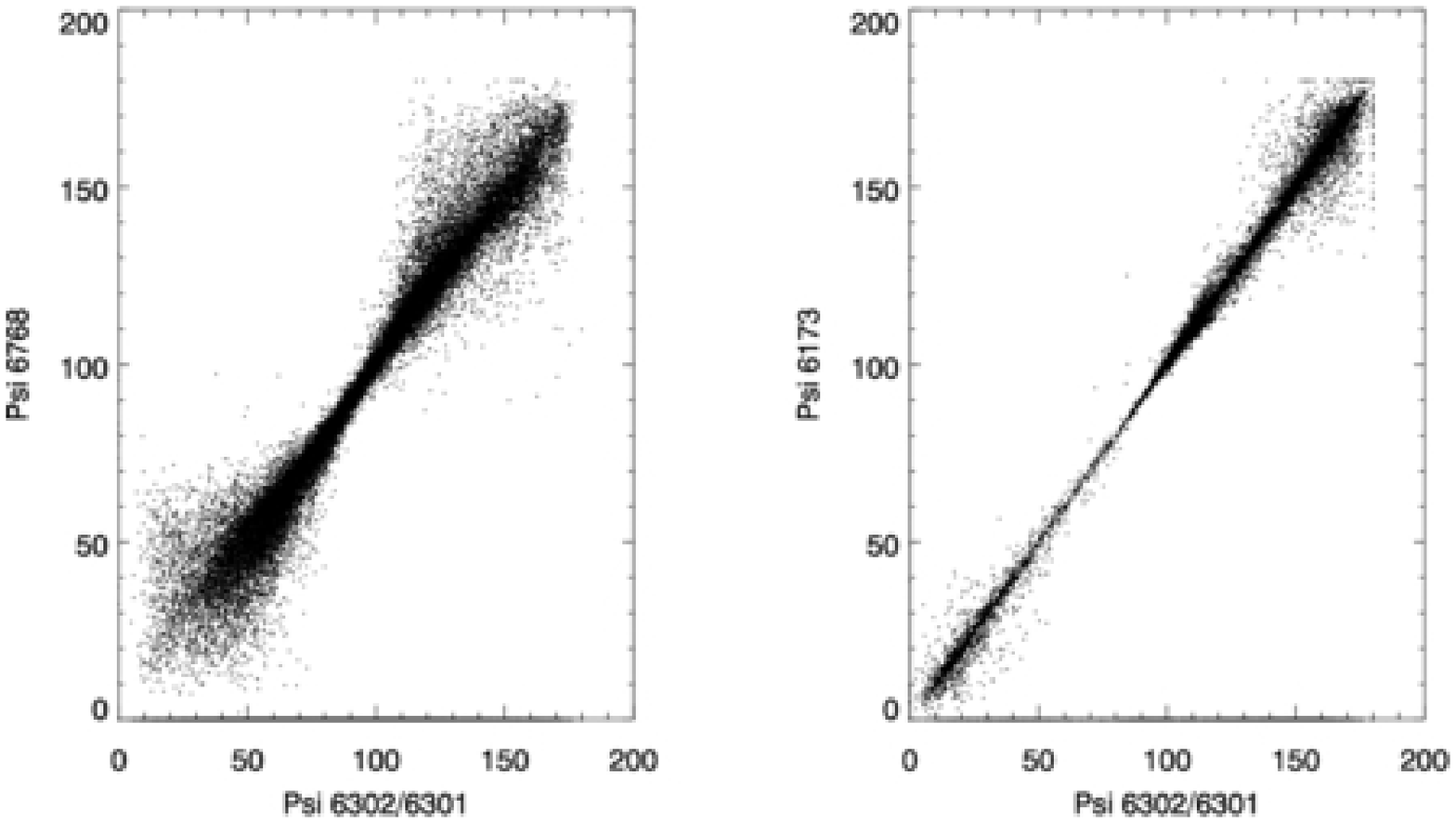,width=25pc}}
\caption{Scatter plots of the values observed simultaneously in different lines.  The left column contains values observed simultaneously with  \ion{Ni}{I} 6768 \AA~and the \ion{Fe}{I} 6302 \AA/6301 \AA~line pair on 2002 March 10.  The right column contains values observed simultaneously with \ion{Fe}{I} 6173 \AA~and the \ion{Fe}{I} 6302 \AA/6301 \AA~line pair on 2002 March 9.  Field strength (top row), filling fraction (middle row - FF)) and inclination (bottom row-Psi) are shown.  }
\label{aspff}
\end{figure*}

Figure~\ref{aspff} shows the scatter plots of the values observed with \ion{Ni}{I} and \ion{Fe}{I} simultaneously with the \ion{Fe}{I} 6302 and 6301 \AA~line pair.  If we assume that the values determined from the dual line inversion of \ion{Fe}{I} 6302 and 6301 \AA~line pair are the ``true" values, then it is obvious from Figure~\ref{aspff} that \ion{Fe}{I} performs better as a vector field diagnostic than \ion{Ni}{I}. This comparison indicates that the higher Land\'e factor, g$_{eff}$, of \ion{Fe}{I} enables a much better determination of the vector magnetic field and filling fraction. 

The increasing uncertainties found in inverted quantities when utilizing a single spectral line for inversion, as opposed to two spectral lines, deserves some comment.  When comparing correlation plots of single-line inversion results $vs$ two-line results, we see larger discrepancies for single line results with a lower g$_{eff}$ (1.4 for 6768) than for a single line with larger g$_{eff}$ (2.5 for 6173). This highlights the necessity for any single line inversion instrument to use a line with a large g$_{eff}$.

\section{Line Performance Using Simulated Profiles}

To estimate line performance, we apply spectral filters (Figure \ref{JGFILTERS}) to simulated line profiles.  The parameters recovered from an ASP observation (discussed in \S3) were used as input to a ME line
profile synthesis code to generate a set of simulated Stokes $I$,  $Q$, $U$ and $V$ profiles for the three spectral lines of \ion{Fe}{I} 6173 \AA, \ion{Fe}{I} 6302 \AA~and \ion{Ni}{I} 6768 \AA.  Filter widths and spacing are scaled by the relative line-center wavelengths to obtain an unbiased comparison.  
We employ only inversion results (the ME parameters) for which the same pixel was inverted in the two observed wavelength regions providing more than 34\,000 sets of ME parameters for the March 9 data and nearly  22\,000 sets of ME parameters for the March 10 data.

\begin{figure*}[htbp]
\centering{\epsfig{file=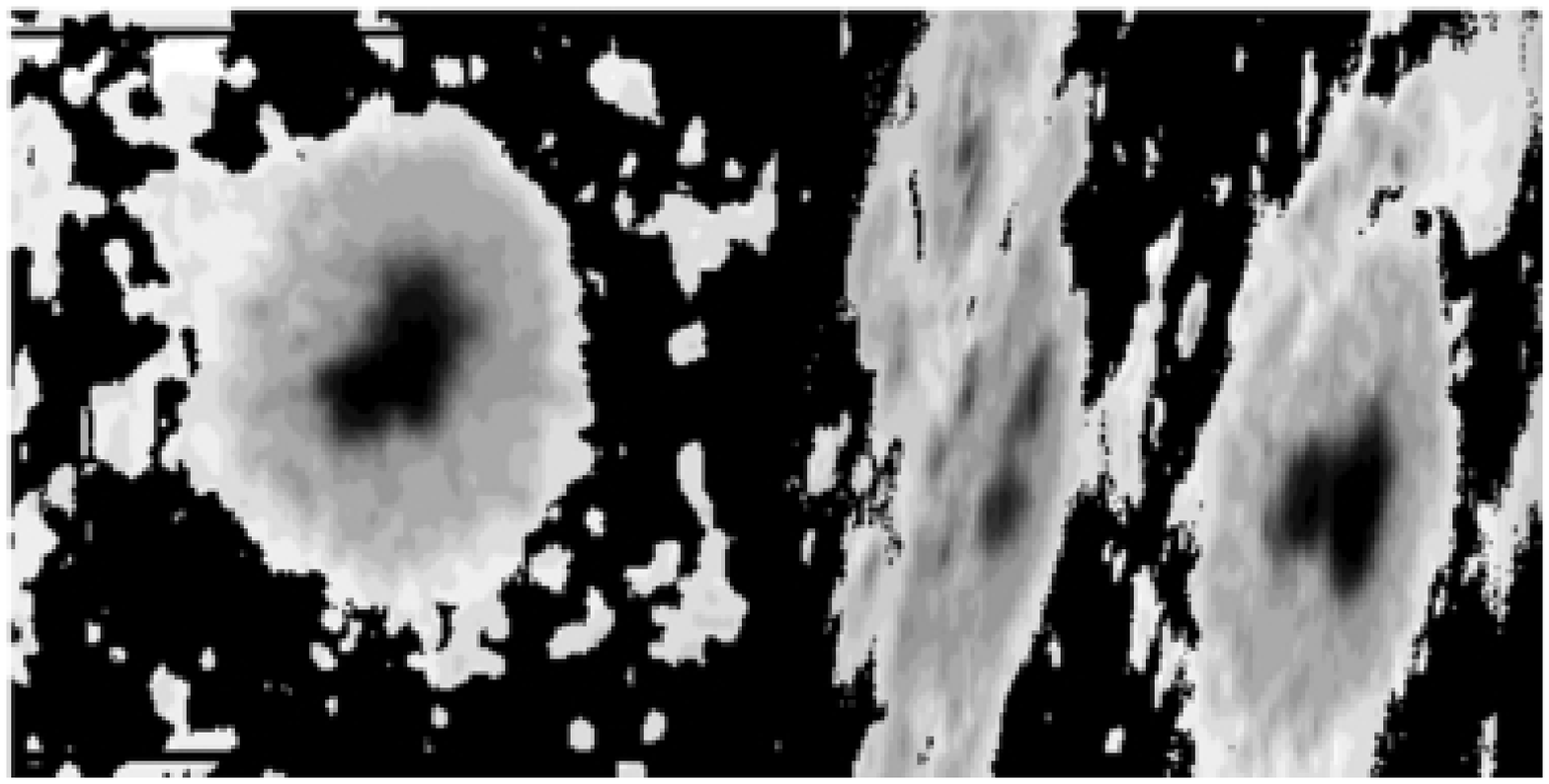,width=25pc}}
\caption{ASP maps for the continuum at 6302~\AA.  The left map is of NOAA 9856,
located at S3 W4, and was made at 19:27 UT on 2002 March 9.  The right map is
of NOAA 9866, located at S9 E62, and was made at 18:58 UT on 2002 March 10.
The black regions outside of the umbra indicate map positions for which inversions were not made.}
\label{JGMAPS}
\end{figure*}

The simulated profiles mimic the observed profiles but lack certain solar affects such as molecular blends in the umbra, nearby lines, and asymmetries caused by gradients in the solar atmosphere.  We have considered that only a fraction, $f$, of our resolution element is occupied by the constant magnetic field (this also accounts for scattered light in the instrument).  This filling factor, $f$, the observed continuum level, and the continuum of the scattered light are all provided by the ASP inversion.  While this approach mimics the observed profiles in most instances, it does not account for other possible distributions of magnetic field inside our resolution element (Emonet and Cattaneo, 2001; Socas-Navarro and Sanchez Almeida, 2003).  Nor do we explore the effects of atmospheres more complicated than ME.  However, Westendorp Plaza {\it et al.} (1998) shows that ME inversions often provide a reasonable mean value of the magnetic field over the line formation region when the conditions for the approximation are not satisfied.  


To better test performance over the expected range of HMI observations, we assign each pixel a new random velocity in the range $\pm$ 6~km/s and random magnetic field orientations.  From Figure~\ref{aspff}, we see that \ion{Ni}{I} appears to be observing a different magnetic field than \ion{Fe}{I}.  Overall, the \ion{Ni}{I} parameters have a lower flux density than the \ion{Fe}{I} parameters.  To remove this bias, we assign the field strength and filling factor inverted for \ion{Fe}{I} 6302~\AA/6301~\AA~to each line that we simulate.  While this gives a more equitable parameter set for comparisons, it may somewhat uncorrelate solar connections between magnetic field and temperature (which is reflected in the ME thermodynamic parameters).  Also, the ASP only inverts when the polarization signal is above a threshold of 4 $\times$ 10$^{-3}$ of the continuum intensity. Therefore, it supplies very few realizations of quiet Sun models.
 
Note that we are not creating an inversion technique. Instead, we analyze the information content of the measurements \todash{}the best possible accuracy that any inversion technique would be able to obtain with HMI filter-polarimetric measurements. To this end, we apply the HMI filters to each ME simulated Stokes profile  and add anticipated HMI noise based on our assumption of measuring $1.5\times10^5$ CCD electrons for the quiet Sun continuum. This noise level means that we can recover our polarization signal with a sensitivity of $2.2\times10^{-3}$ of the intensity continuum.  This is defined as a simulated observation.

We employ a weighted Levenberg-Marquardt (LM) least-squares minimization to fit Milne-Eddington models to our simulated observation (del Toro Iniesta and Ruiz Cobo, 1996). By using the pre-noise input ME parameters as initial guesses, we seek to determine how much information was lost by the application of filters and noise.  The spectral sampling limitation combined with the noise may cause the fit to not occur precisely at the input parameters.  There may be no well-defined minimum.  Instead, a range of parameters may all yield the same minimum $\chi^2$ value.
The statistical difference between the fitted and the true parameters provides a measure of the information content contained in the filtered spectral line.  The results are verified by setting a new initial guess (original ME parameters $+$ $10$~m/s, $50~G$, $2^\circ$ in each $\psi$ and $\phi$, and 0.05 in $f$) and running a second minimization.  The results are essentially unchanged, confirming that our method is not sensitive to small changes in initialization.

 We supply the fitting code with both the scattered-light profile and the continuum level for the simulated observation which we assume to be previously obtained.  It is not clear at this time how these quantities will be obtained for HMI.  It is clear, however, that they are essential for the accurate determination of the filling factor and intrinsic magnetic field strength.

\subsection{Line Comparisons}

\begin{figure*}[htbp]
\centering{\epsfig{file=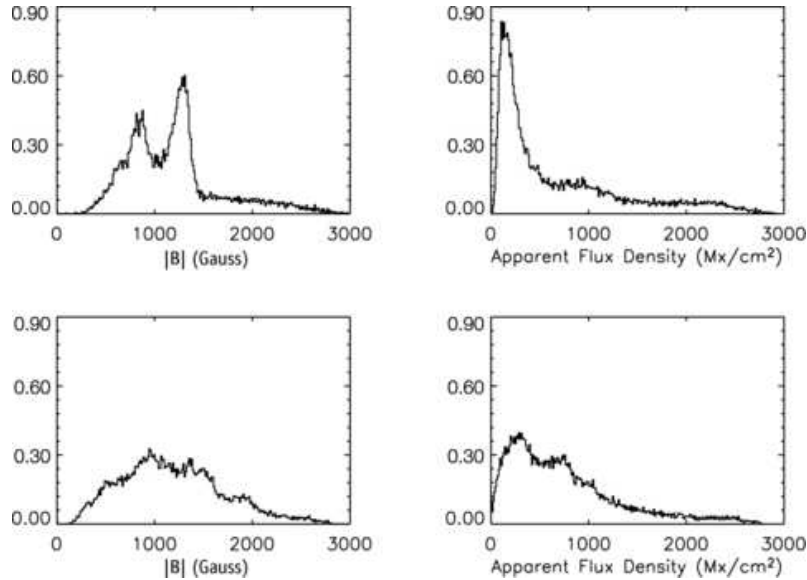,width=25pc}}
\caption{Histograms of ASP inverted ME parameters from spectrally resolved observations of \ion{Fe}{I} 6302~\AA/6301~\AA.  The top panels are for NOAA 9856 and the lower panels are for NOAA 9866.}
\label{JGPARAMS}
\end{figure*}

\begin{figure*}[htbp]
\centering{\epsfig{file=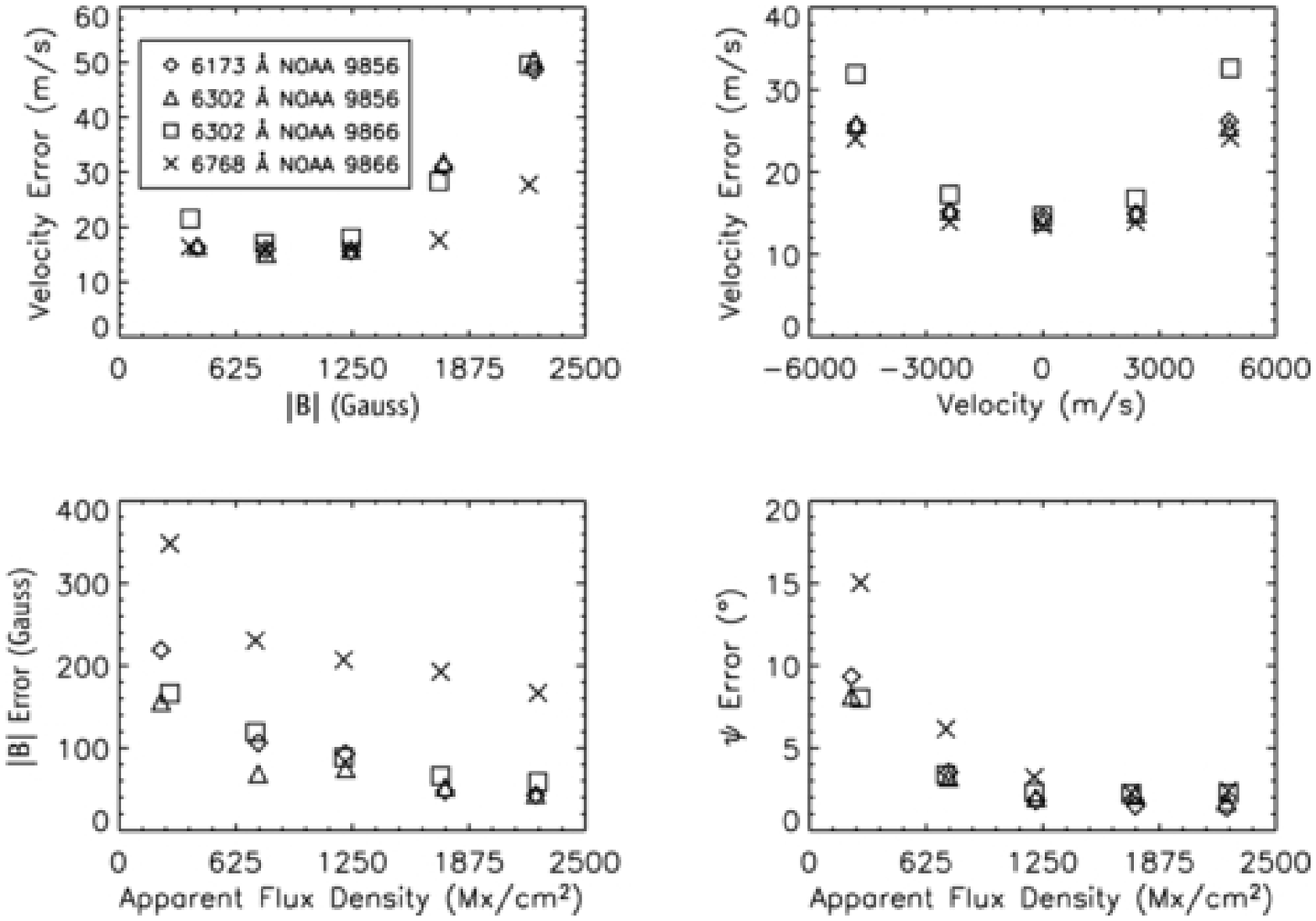,width=25pc}}
\centering{\epsfig{file=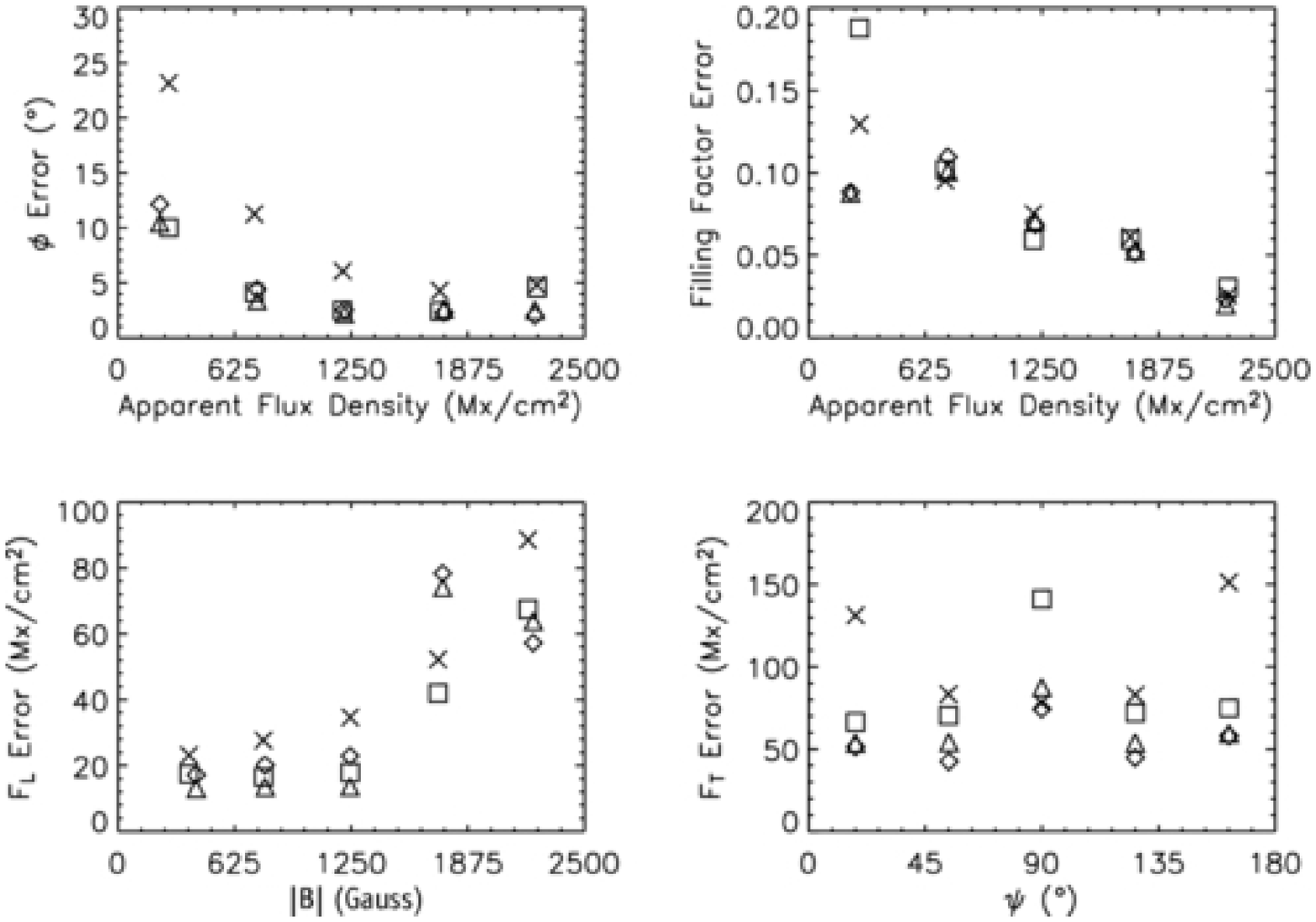,width=25pc}}
\caption{Spectral-line performance comparison using simulated profiles for \ion{Fe}{I} 6173~\AA~from NOAA 9856 ($\diamond$), \ion{Fe}{I} 6302~\AA~from NOAA 9856 ($\triangle$),  \ion{Fe}{I} 6302~\AA~from NOAA 9866 (box), and \ion{Ni}{I} 6768~\AA~from NOAA 9866 ($\times$).  Velocity errors are plotted $vs$ $|\bf{B}|$ (top left) and velocity (top right).  Errors of $|\bf{B}|$ (second row left) and inclination (second row right) are plotted $vs$ apparent flux density.  Azimuth (third row left) and $f$ (third row right) are plotted $vs$ apparent flux density.  $F_L$ $vs$ $|\bf{B}|$ (lower left) and for $F_T$ $vs$ inclination are shown. These results represent the 68\% confidence level.}
\label{JGSIM1}
\end{figure*}

Before comparing the spectral line performance, note the very different distributions of the magnetic field strength, $|\bf{B}|$, apparent flux density, $F = f \times |\bf{B}|$  (see Figure \ref{JGPARAMS}), line depth and width (see Figure~\ref{cl6768}) for the two active regions.  While some difference between any two active regions is to be expected, much of the distribution difference seen here is due to the different center-to-limb angles at which they were observed.  Performance comparisons for the simulated observations are found in Figure \ref{JGSIM1}.   

In Figure \ref{JGSIM1}, note that all three lines have similar velocity errors for weak fields and small input velocities 
The ``saturation" effect discussed in \S 2.2 is seen in Figure \ref{JGSIM1} as the effective velocity range is smaller for the \ion{Fe}{I} lines than for the \ion{Ni}{I} line in the umbra (additionally, the filter set is 10\% wider for \ion{Ni}{I} 6768~\AA~than for \ion{Fe}{I} 6173~\AA).  

For magnetic field strength and inclination to the line of sight there is nearly a factor of two better performance for the \ion{Fe}{I} lines than the \ion{Ni}{I} line. Errors decrease as the apparent flux density, and therefore polarization signal strength, increases.  These results reveal an important tradeoff.  While \ion{Fe}{I} has a better magnetic performance by a factor of two than \ion{Ni}{I} and similar velocity performance for weak fields, it will be necessary to ``chase the line'' during spacecraft motion or suffer a less accurate velocity performance in umbrae. 

Note that the \ion{Fe}{I} 6302~\AA~performance at disk center is slightly better than for the same line at $\mu = 0.6$ and this should be kept in mind when comparing \ion{Fe}{I} 6173~\AA~(observed at $\mu = 1$) and \ion{Ni}{I} 6768~\AA~(observed at $\mu = 0.6$). Even so, the performance difference is very small in most cases and allows for an almost direct comparison between all three lines.

In Figure \ref{JGSIM1}, note that while all lines have azimuth errors less than five degrees for strong polarization signals, \ion{Ni}{I} is noticeably less accurate than \ion{Fe}{I} for low apparent-flux density.  $f$ errors decrease for larger polarization signals, as expected.   For low field strengths, there are relatively low errors for apparent longitudinal flux density, $F_L = f \times |\bf{B}| \times \cos{\psi}$. 
In the transition between the weak and strong-field regimes, the Stokes $V$ signal becomes more complicated.  Once the line is fully split, interpretation becomes simpler and, indeed, the errors for $F_L$ decrease for the \ion{Fe}{I} lines for the strongest fields we examined. 

The errors shown in Figure 8 are the absolute values of the errors.  If the absolute values were not taken, then the errors should be near zero unless systematic effects are present.  We do not show a separate plot for the systematic errors, but want to mention them.   The systematic errors for velocity determination are less than $1$~m/s for all lines in the range of 0 \todash $\pm$3 km/s.  For the extremes of the velocity range, the systematic errors underestimate the magnitude of the velocity by about $3$~m/s for all lines.The magnetic field strength as plotted $vs$ flux density (similar to second row, left plot of Figure~\ref{JGSIM1} except systematic errors would be plotted about the abscissa) shows a systematic overestimation for the \ion{Ni}{I} line of $150$~G and a systematic underestimation for the \ion{Fe}{I} lines of $50$~G at $250$~Mx/cm$^2$. These errors reverse sign above $2$~kMx/cm$^2$ and decrease in magnitude to $\pm20$~G.   The filling factor errors include a systematic overestimation of $f$ for all lines of about half the magnitude of what is plotted in the third row, right column of Figure~\ref{JGSIM1}.  The \ion{Ni}{I} line shows no appreciable systematic errors for $F_L$ while the \ion{Fe}{I} lines show a systematic underestimation by one or two Mx/cm$^2$ for both very weak and very strong fields.  These were the only noticeable systematic errors.  

Regarding $F_T = f \times |\bf{B}| \times \sin{\psi}$, there is some curious behavior.  The \ion{Fe}{I} lines, especially 6302~\AA~from NOAA 9866,  have larger errors at $90^\circ$ while \ion{Ni}{I} errors are smaller. Smaller errors are expected as the Stokes $Q$ and $U$ signals will be stronger at $90^\circ$ than at other angles.   We hypothesize that this lower performance is due in part to the lower signal to noise ratio in general for Stokes $Q$ and $U$ as opposed to Stokes $V$ and to the relatively low spectral resolution of HMI.  This hypothesis is supported by the observation that $F_L$, which is mostly determined from Stokes $V$, is better determined by nearly a factor of two.
Alternatively, we suggest that $F_T$ may be poorly determined for each line for transverse fields of certain magnetic field strengths especially near the limb.  Since the poor performance is more striking for the NOAA 9866 \ion{Fe}{I} 6302~\AA~data than for the NOAA 9856 \ion{Fe}{I} 6302~\AA~data, this may have something to do with line depths and widths due to the differing distributions of magnetic field strengths shown in Figure \ref{JGPARAMS}.  

\subsection{Magnetograph Mode}

HMI will utilize two CCD cameras.   One camera, the ``Doppler" camera, will be observing only Stokes $I$ and $V$ while running at a higher cadence than the vector polarimetry camera.  To make performance estimations for this mode of operation, we repeat the numerical experiments of the previous subsection only using Stokes $I$ and $V$ information. The LM fits are initialized by the pre-noise input parameters altered in the following way:  the azimuth is set to zero and not fitted, the inclination is set to $2^\circ$ and $|\bf{B}|$ is projected onto this new angle.  For verification that our results reflect  the correct $\chi^2$ minima, we re-run the fits with the inclination set to a random value and $|\bf{B}|$ is projected correspondingly.  This is done because our LM code fits separately for $|\bf{B}|$, $f$, and $\psi$ instead of simply for $F_L$. Though not fitting for $F_L$ directly may impact the applicability of our technique,  the results we obtain from these two fittings are essentially unchanged, confirming that we have found the correct $\chi^2$ minimum.

\begin{figure*}[htbp]
\centering{\epsfig{file=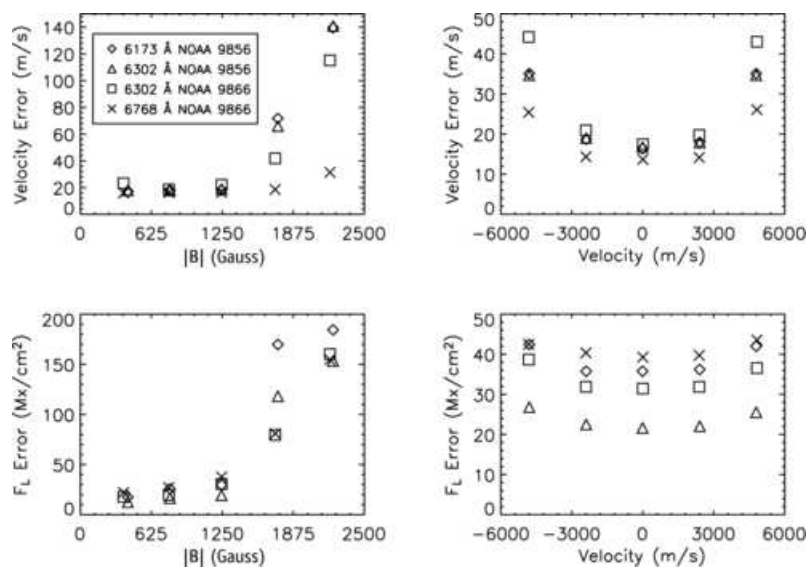,width=25pc}}
\caption{Performance comparison using only simulated Stokes $I$ and $V$ for \ion{Fe}{I} 6173~\AA~from NOAA 9856 ($\diamond$), \ion{Fe}{I}6302~\AA~from NOAA 9856 ($\triangle$),  \ion{Fe}{I} 6302~\AA~from NOAA 9866 (box), and \ion{Ni}{I} 6768~\AA~from NOAA 9866 ($\times$).  The top panels are velocity errors $vs$ $|\bf{B}|$ (left), velocity (right).  The bottom panels are errors of $F_L$ $vs$ $|\bf{B}|$ (left), velocity (right). The results represent the 68\% confidence level.}
\label{JGSIMIV}
\end{figure*}

In Figure \ref{JGSIMIV}, we find the following differences in line performance for the magnetograph mode compared to when the full Stokes vector is observed.  Velocity errors are subject to the same ``saturation" effect, though now all errors are increased over their previous levels due to the loss of the information in Stokes $Q$ and $U$.  As before, errors for $F_L$ increase until the lines become fully separated.  It is interesting to note that $F_L$ is determined less precisely without Stokes $Q$ and $U$, even in regions of strong magnetic field.  Because of this, \ion{Fe}{I} lines no longer perform better than \ion{Ni}{I} for umbral $F_L$
measurements.  For very low field strengths, the \ion{Fe}{I} lines perform only about $10$~Mx/cm$^2$ better than the \ion{Ni}{I} line.  The performance for $F_L$ is also a stronger function of velocity for the \ion{Fe}{I} lines than for \ion{Ni}{I}.  The trade-off for magnetograph mode is less pronounced than for the vector polarimetry mode.  \ion{Ni}{I} will better measure velocity in umbra while the \ion{Fe}{I} will measure $F_L$ better in regions of weak fields.  We conclude that the full benefit of using the \ion{Fe}{I} line only occurs when full vector polarimetry is utilized.

\section{Line Performance Using Observed Profiles}

As in Section 4, we apply simulated spectral filters (Figure \ref{JGFILTERS}) to the observed line profiles and apply noise such that we can recover polarization with a precision of $2.2\times10^{-3}$.  The calculations are carried out identically to those of Section 4 except that observed profiles are used in place of simulated profiles.  \ion{Fe}{I} 6301~\AA~/ 6302~\AA~data has telluric lines nearby and therefore were not analyzed by applying filters.   ASP inversion results are used as the initial guess for the LM fit.  These results using observed profiles, then, provide an indication of the attainable error levels for HMI with the proposed filters and polarization precision.

To estimate errors in the LM fits to the filtered observed profile data, we assume that the ASP results are the ``true values" and the errors are the difference between the LM fits and the ASP inversions.  The true  performance of the observed data lies somewhere between the quoted value and that value minus the uncertainty in the ASP inversion.  Note that the velocities employed in these calculations are not redistributed randomly as done with the simulated profile data.  Instead, the velocities are the real solar velocities as observed with ASP.  
In addition to the differences between the active regions (see Figure \ref{JGPARAMS}), NOAA 9856 for \ion{Fe}{I} 6173~\AA~ and NOAA 9866 for \ion{Ni}{I} 6768~\AA, it must be kept in mind that the \ion{Fe}{I} line was observed at disk center and the \ion{Ni}{I} line was observed at $\mu = 0.6$.

\subsection{\ion{Fe}{I} 6173~\AA}

\begin{figure*}[htbp]
\centering{\epsfig{file=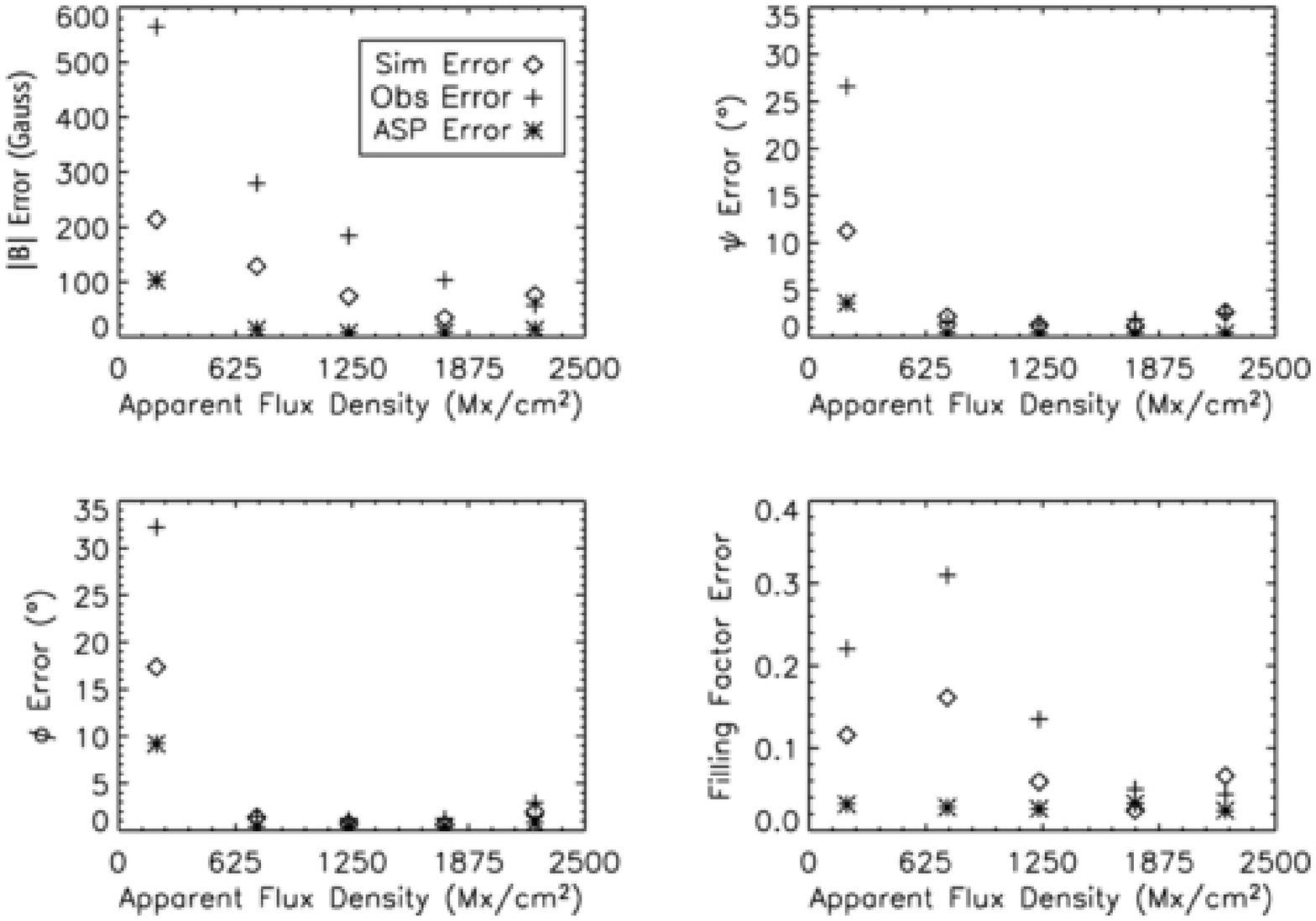,width=25pc}}
\centering{\epsfig{file=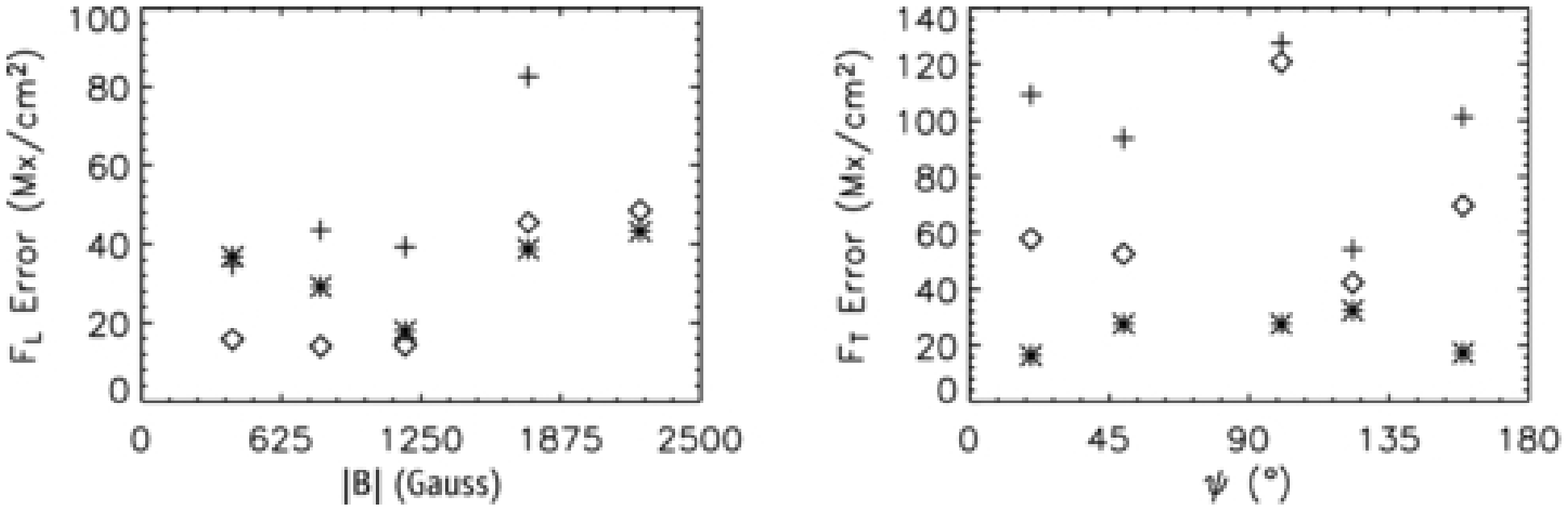, width=25pc}}
\caption{68\% confidence intervals for \ion{Fe}{I} 6173~\AA.  Errors are shown for the ASP \ion{Fe}{I} inversion data (*),  the observed profile data (+) and the simulated profile data ($\diamond$).  Field strength (upper left), inclination (upper right), azimuth (middle left), and filling factor (middle right) are all plotted $vs$ apparent flux density. $F_L$ errors are shown $vs$ magnetic field strength (lower left) and $F_T$ errors are shown $vs$ inclination (lower right).  (Note that the data points for $F_L$ for  the observed profiles and ASP uncertainty coincide at $|\bf{B}| =$ 2250~G.)}
\label{JGASP1}
\end{figure*}

As seen in Figure \ref{JGASP1}, the observed profiles performance is poor for apparent flux density near $250$~Mx/cm$^2$.  Above $600$~Mx/cm$^2$, however, field direction (inclination and azimuth) can be determined to within a few degrees.  For apparent flux density above $1500$~Mx/cm$^2$, magnetic field strength can be determined to within $100$~G. The observations are less accurate than the simulations, but the performance does improve as the polarization signal increases.  

$F_L$ and $F_T$ have the same dependencies on magnetic field strength and inclination as discussed for the simulations in Section 4.  In general, the accuracy will be about $40~$Mx/cm$^2$ for $F_L$ and $100$~Mx/cm$^2$ for $F_T$.   It appears that the observed profile data performance for \ion{Fe}{I} should be reasonable for $F_L$ and $F_T$ in general, for field directions in all regions with more than $600$~Mx/cm$^2$, and for magnetic field strength in penumbra and umbra.

\subsection{\ion{Ni}{I} 6768~\AA}

\begin{figure*}[htbp]
\centering{\epsfig{file=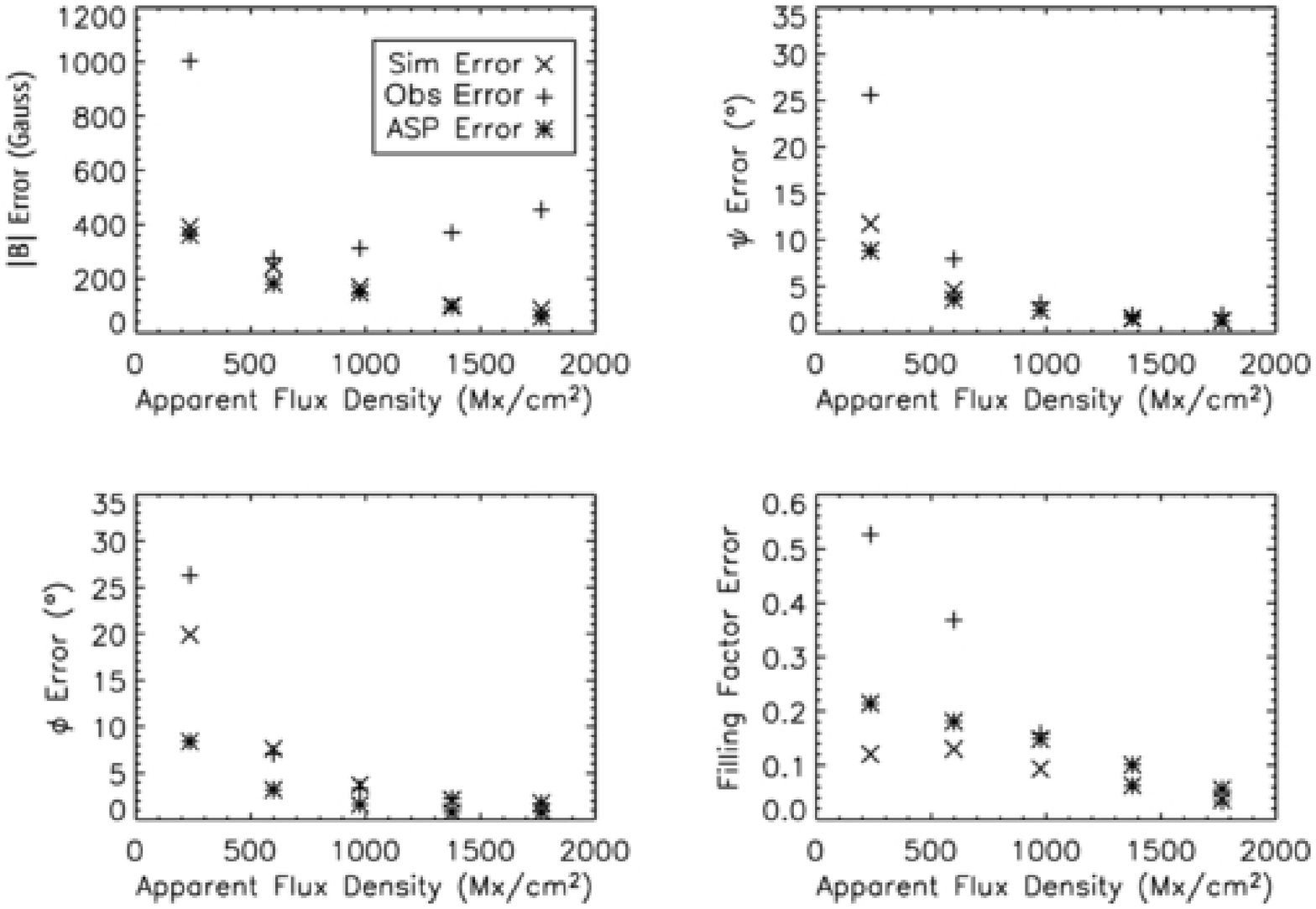,width=25pc}}
\centering{\epsfig{file=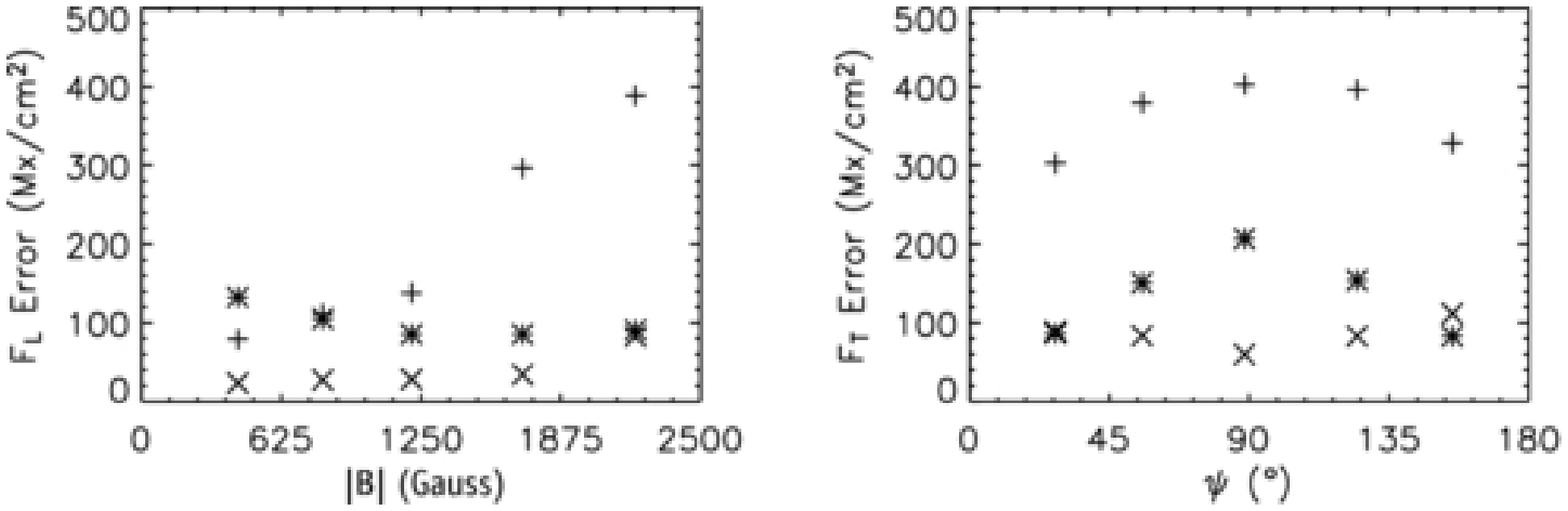,width=24pc}}
\caption{68\% confidence intervals for \ion{Ni}{I} 6768~\AA.  Errors are shown for the ASP \ion{Fe}{I} inversion data (*),  the observed profile data (+) and the simulated profile data ($\times$).  Field strength (upper left), inclination (upper right), azimuth (middle left), and filling factor (middle right) are all plotted $vs$ apparent flux density. $F_L$ errors are shown $vs$ magnetic field strength (lower left) and $F_T$ errors are shown $vs$ inclination (lower right).}
\label{JGASP3}
\end{figure*}

As seen in Figure \ref{JGASP3}, ASP uncertainties for magnetic field strength for this line are a factor of four greater for the \ion{Fe}{I} line.  Oddly, the simulated errors are generally as good or better than ASP.  This is not the situation for \ion{Fe}{I} and may indicate that uncertainties are coming from the line itself when used for vector field determination rather than the spectral filtering of the data.  Indeed, the ASP ME fits to the observed profiles are of poorer quality for \ion{Ni}{I}. There are significant asymmetries outside of the umbra and umbral contamination is more troublesome than for \ion{Fe}{I}.  These problems appear to manifest themselves in the magnetic field strength observed profiles at  $1$~kMx/cm$^2$ and above.

Nevertheless, field directions are generally recovered as well for this line as for \ion{Fe}{I}, but not until $1$~kMx/cm$^2$. The difficulties for magnetic field strength are also experienced for $F_L$ and $F_T$ as would be expected.  It is interesting to note that the poor $F_T$ performance at $90^\circ$ exhibits itself for ASP uncertainties for both \ion{Ni}{I} and \ion{Fe}{I}.  This agrees with our hypothesis from Section 4 that the lower signal-to-noise in Stokes $Q$ and $U$ as opposed to Stokes $V$ is partially responsible for these errors.  From this data, we only predict acceptable performance for \ion{Fe}{I} 6173~\AA.

\section{Conclusions}

Both \ion{Fe}{I} 6173 \AA~and \ion{Ni}{I} 6768 \AA~profiles have clean continuum and no blends which threaten performance (see Figures~\ref{6173w} and~\ref{6173asp}).  The \ion{Ni}{I} parameters of line width, line depth and wing slope show smooth and predictable changes as a function of center-to-limb angle (see Figure~\ref{cl6768}). \ion{Ni}{I} has been carefully studied by Bruls (1990) who concluded the line was a good choice for helioseismology because the line profile is stable and not very sensitive to variation of temperature and temperature gradient in the photosphere.  Extensive studies of this nature, including effects of granulation on the line formation, are yet to be carried out for \ion{Fe}{I}. 

Not surprisingly, the higher $g_{eff}$ \ion{Fe}{I} serves as a much better diagnostic tool than \ion{Ni}{I} to ascertain the inclination angle, field strength and filling factor of the magnetic field.  This is a robust result  as it is obtained repeatedly using a variety of analysis techniques.  First, the ASP observations of \ion{Fe}{I} and \ion{Ni}{I} compared to the ASP observations of the benchmark  \ion{Fe}{I} 6301 \AA~/ 6301 \AA~ line pair show \ion{Ni}{I} to have much greater uncertainty in recovering the field strength, filling factor and inclination  (see Figure~\ref{aspff}).  

Second, the simulated line profile data shows \ion{Fe}{I} to perform nearly a factor of two better than the \ion{Ni}{I} line for field strength and inclination determination (see Figure~\ref{JGSIM1}).  Third, results using observed profiles show that \ion{Fe}{I} is able to determine field strength, longitudinal and transverse flux four times more accurately than \ion{Ni}{I} in active regions.    Inversions using observed profiles show inclination and azimuthal errors to be recovered to $\approx$2$^\circ$ degrees above 600 Mx/cm$^2$ for \ion{Fe}{I} and above 1000 Mx/cm$^2$ for \ion{Ni}{I} (see Figures~\ref{JGASP1} and \ref{JGASP3}).  This means, using the \ion{Fe}{I} line allows better magnetic field orientation determination in plage, whereas both lines will provide a good orientation determination in penumbra and umbra.    From the above-mentioned results, we predict acceptable performance for \ion{Fe}{I} 6173 \AA~for vector magnetic field determination.  

In recovering velocities, the performance of the spectral lines from inversion methods shows both the \ion{Fe}{I} and \ion{Ni}{I} lines perform well for low field strengths and low velocities, with errors of roughly 15 m/s.  However, the higher $g_{eff}$ of \ion{Fe}{I} means that its operational range of velocity values in regions of strong magnetic field is smaller than \ion{Ni}{I}.  \ion{Fe}{I} performs poorly at high field strengths and high velocities where the line has moved beyond the filter sampling range.  This implies that in order to have accurate velocity measurements in umbrae using the \ion{Fe}{I} line, the line sampling must be increased by either increasing the number of filter positions or chasing the line by changing the filter positions as a function of spacecraft velocity.

Simulations using only the Stokes $I$ and $V$ profiles shows that the velocity errors increase for both spectral lines in umbra if Stokes $Q$ and $U$ are not considered (see Figure~\ref{JGSIMIV}  as compared to Figure~\ref{JGSIM1}) .  This result implies that active region helioseismology accuracy will improve if the full Stokes vector is used for analysis.   

Using the observed profile performance for \ion{Fe}{I} as a good indicator for its performance on-board the HMI spacecraft, we conclude that it will perform well for $F_L$ and $F_T$ in general, for field directions in all regions with more than $600$~Mx/cm$^2$, and for magnetic field strength in penumbra and umbra.


\acknowledgements
    The authors thank the National Solar Observatory for use of the Advanced Stokes Polarimeter.  The authors thank the Helioseismic Magnetic Imager team for support and guidance.  NASA grant monies NAS5-02139 supported much of the research conducted for this paper.  We are grateful to Juan Borrero and John Leibacher for their helpful comments.

\end{article}

\begin{thebibliography}{}

\bibitem[Auer {\it et al.}, (1977)]{aue77}
Auer, L.H. {\it et al.}, 1977, {\it Solar Phys.}, {\bf 55}, 47.

\bibitem[Bruls, (1993)]{bru93}
Bruls, J.H.M.J., 1993, {\it A. Ap.}, {\bf 269}, 509.

\bibitem[del Toro Iniesta {\it et al.}, (1996)]{del96}
del Toro Iniesta, J. C. and B. Ruiz Cobo, 1996, {\it Solar Phys.}, {\bf 164}, 169.

\bibitem[Emonet and Cattaneo, (2001)]{emo01}
Emonet, T.  and  F. Cattaneo, F., 2001, {\it Astrophys. J.}, {\bf 560}, L197.

\bibitem[Graham {\it et al.}, (2002)]{gra02}
Graham, J.D., L\'opez Ariste, A., Socas-Navarro, H. and Tomczyk, S., 2002, {\it Solar Phys.}, 2008, 211. 

\bibitem[Landi degl'Innocenti, (1982)]{lan82}
Landi degl'Innocenti, E. 1982, {\it Solar Phys.}, {\bf 77}, 285.

\bibitem[Jones, (1989)]{jon89}
Jones, H., 1989, {\it Solar Phys.}, {\bf 120}, 211.

\bibitem[Maltby {\it et al.}, (1986)]{mal86}
Maltby, P., Avrett, E. H., Carlsson, M., Kjeldseth-Moe, O., Kurucz, R. L., and  
Loeser, R., 1986, {\it Astrophys. J.}, {\bf 306}, 284.

\bibitem[Rybicki and Hummer, (1991)]{ryb91}
Rybicki, G.B and Hummer, D.G. 1991, {\it A. Ap.}, {\bf 245}, 171.

\bibitem[Simmons and Blackwell, (1982)]{sim82}
Simmons, G.J., and D.E. Blackwell, 1982, {\it A. Ap.}, {\bf 112}, 209.

\bibitem[Socas-Navarro and Sanchez Almeida, (2003)]{soc03}
Socas-Navarro, H. and Sanchez Almeida, J., 2003, {\it Astrophys. J.}, {\bf 593}, 581.

\bibitem[Solanki and Stenflo, (1985)]{sol85}
Solanki, S.K. and J.O. Stenflo, 1985, {\it A. Ap.}, {\bf 148}, 123.

\bibitem[Stenflo and Lindegren, (1977)]{ste77}
Stenflo, J.O. and Lindegren, L., 1977, {\it A. Ap.}, {\bf 59}, 367.

\bibitem[Uitenbroek, (2001)]{uit01}
Uitenbroek, H., 2001, {\it Ap. J.}, {\bf 557}, 389.

\bibitem[Ulrich{\it et al.}, (1991)]{ulr91}
Ulrich, R.K., Webster, L., Boyden, J.E., Magnone, N., and Bogart, R.S., 1991, {\it Solar Phys.}, {\bf 135}, 211.  

\bibitem[Ulrich{\it et al.}, (2002)]{ulr02}
Ulrich, R.K., Evans, S., Boyden, J.E., and Webster, L., 2002, {\it Astrophys. J.S.}, {\bf 139}, 259.

\bibitem[Vernazza, {\it et al.} (1973)]{ver73}
Vernazza, J.E., Avrett, E.H. and Loeser, R.: 1973, {\it Astrophys. J.}, {\bf 184}, 605.

\bibitem[Vernazza, {\it et al.} (1976)]{ver76}
Vernazza, J.E., Avrett, E.H., and Loeser, R.: 1976, {\it Astrophys. J.S.}, {\bf 30}, 1.

\bibitem[Vernazza, {\it et al.} (1981)]{ver81}
Vernazza, J.E., Avrett, E.H., and Loeser, R.: 1981, {\it Astrophys. J.S.}, {\bf 45}, 635.

\bibitem[Westendorp Plaza, {\it et al.} (1998)]{wes98}
Westendorp Plaza, C., del Toro Iniesta, J.C., Ruiz Cobo, B., Martinez Pillet, V., Lites, B.W. and Skumanich, A., 1998, {\it Astrophys. J.}, {\bf 494}, 453.
\end{thebibliography}
\end{document}